\documentclass[aps,prl,reprint,superscriptaddress,citesort,nolongbibliography]{revtex4-2}
\usepackage{graphicx}
\usepackage{amsmath, amssymb}

\usepackage{tabularx}
\usepackage{csquotes}

\usepackage{booktabs,threeparttable}
\usepackage{mhchem}
\usepackage{bm}

\usepackage{hyperref}
\usepackage{xcolor}
\definecolor{tab_blue}{HTML}{1F77B4}
\hypersetup{%
  breaklinks=true,
  colorlinks=true,
  linkcolor=tab_blue,
  filecolor=tab_blue,
  urlcolor=tab_blue,
  citecolor=tab_blue,
}

\def\be{\begin{equation}}
\def\ee{\end{equation}}
\def\ba{\begin{eqnarray}}
\def\ea{\end{eqnarray}}
\def\epa{\boldsymbol{e}_\parallel}
\def\kpn{k_{\perp}}
\def\ppn{p_{\perp}}
\def\qpn{q_{\perp}}
\def\kpa{k_{\parallel}}
\def\kk{\boldsymbol{k}}
\def\pp{\boldsymbol{p}}
\def\qq{\boldsymbol{q}}
\def\kknb{\boldsymbol{k_{\perp}}}
\def\ppnb{\boldsymbol{p_{\perp}}}
\def\qqnb{\boldsymbol{q_{\perp}}}
\def\uu{{\boldsymbol u}}

\def\bnabla{{\boldsymbol \nabla}}

\begin{document}

\title{Odd wave turbulence}
\author{Xander\,M.\,de\,Wit}
\thanks{Both authors contributed equally to this work.}
\affiliation{Department of Applied Physics, Eindhoven University of Technology, 5600 MB Eindhoven, Netherlands}
\author{L\'eo\,Touzo}
\thanks{Both authors contributed equally to this work.}
\affiliation{James Franck Institute, The University of Chicago, Chicago, IL 60637, USA}
\author{Sébastien\,Galtier}
\affiliation{Université Paris-Saclay, Laboratoire de Physique des Plasmas, École polytechnique, 91128 Palaiseau, France}
\author{Michel\,Fruchart}
\affiliation{Gulliver, ESPCI Paris, Université PSL, CNRS, 75005 Paris, France}
\author{Federico\,Toschi}
\affiliation{Department of Applied Physics, Eindhoven University of Technology, 5600 MB Eindhoven, Netherlands}
\affiliation{CNR-IAC, I-00185 Rome, Italy}
\author{Vincenzo\,Vitelli}
\email{vitelli@uchicago.edu}
\affiliation{James Franck Institute, The University of Chicago, Chicago, IL 60637, USA}
\affiliation{Leinweber Institute for Theoretical Physics, The University of Chicago, Chicago, IL 60637, USA}

\date{\today}

\begin{abstract}

Wave turbulence describes the statistical dynamics of weakly interacting waves in out-of-equilibrium systems. 
These systems are out of equilibrium because forcing and dissipation occur at different scales, leading to a flux across scales known as a turbulent cascade. 
However, the underlying medium is usually assumed to be passive.
Here, we study situations where a turbulent cascade takes place on top of a driven-dissipative steady-state, which is already out of equilibrium without forcing.
We focus on chiral active media, in which waves arise as a consequence of non-reciprocal responses known as odd viscosity and odd elasticity. 
Combining analytical theory with large-scale direct numerical simulations, we show that fluids with odd viscosity sustain weak wave turbulence down to the smallest active scales, leading to anisotropic Kolmogorov–Zakharov spectra and a modified forward cascade.
In the odd elastic solids, two new conserved quantities emerge (which we call wave action and odd energy), giving rise to an inverse cascade in the weak regime and a forward cascade in the strong regime.
Our results pave the way towards understanding the chaotic nonlinear regime of systems with nonreciprocal responses from chiral plasma to engineered solids.

\end{abstract}

\maketitle	

Wave turbulence describes the long-time statistical behavior of out-of-equilibrium systems composed of weakly interacting waves \cite{Zakharov1992,Nazarenko11,GaltierCUP2023,Newell2011,Cosmo2017,Zhou2021}. 
It has been applied across scales from acoustic waves in elastic sheets to gravitational waves.
In these situations, driving and dissipation are negligible in the so-called inertial range of scales where a turbulent cascade takes place.
In this Letter, we combine exact analytical calculations and large scale direct numerical simulations to elucidate how wave turbulence occurs in nonequilibrium media activated at microscopic scales.
We focus on two prototypical examples of chiral waves in active media: fluids with odd viscosity and solids with odd elasticity \cite{oddreview}.
We dub this realm {\it odd wave turbulence}.
Both odd viscosity and elasticity arise in microscopic systems driven out of equilibrium, but at the macroscopic level they have opposite properties with respect to energy dissipation: odd viscosity conserves energy, while odd elasticity corresponds to energy gain and loss. 
As a consequence, odd viscous fluids exhibit a modified version of the energy cascade present in standard fluids.
In contrast, elastic energy is not conserved in odd elastic solids; instead, the transfer across scales is governed by a new emergent conserved quantity, which we refer to as the wave action.

\begin{figure*}
    \centering
    \includegraphics[width=0.995\linewidth]{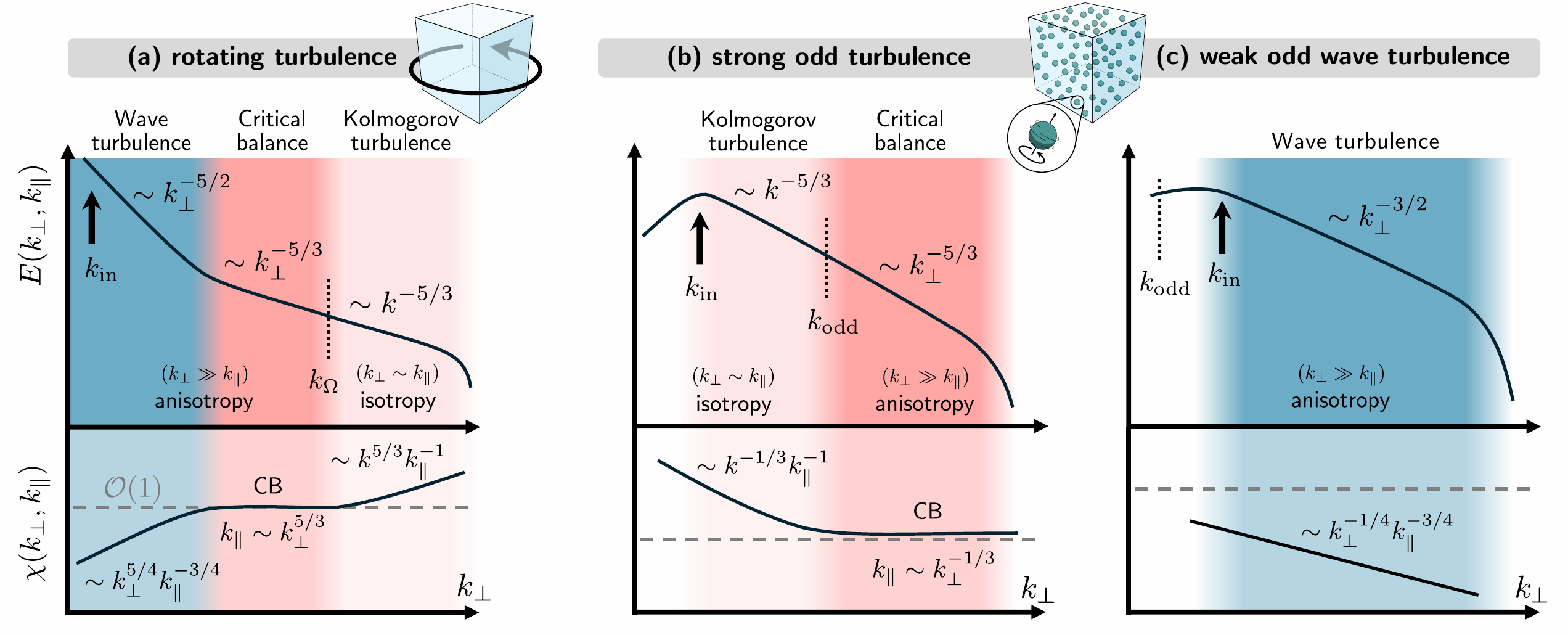}
    \caption{\textbf{Wave turbulence in rotating fluids and in fluids with odd viscosity.}
    The top row depicts the anisotropic energy spectra $E(k_\perp, k_\parallel)$, while the bottom row depicts the time scale ratio $\chi\equiv\tau_{\textrm{lin}} / \tau_{\textrm{NL}}$ for some $k_\parallel>0$. The dashed line in the bottom panel denotes the critical value $\chi\approx\mathcal{O}(1)$ below which wave turbulence is expected. For rotating wave turbulence (a), the dynamics gradually becomes less weak and ultimately recovers strong turbulence and isotropization at small scales beyond the Zeman wavenumber $k_\Omega$~\cite{Zhou1995,Zeman1994,Alexakis2018,Zhou2021}. By contrast, in odd turbulence, in the range $k>k_\textrm{in}>k_\textrm{odd}$ (b), weak wave turbulence is sustained all the way down to the smallest active scales, while strong turbulence is only obtained if energy is injected at large scales $k_\textrm{in}<k_\textrm{odd}$ (c).
    }
    \label{fig:intro}
\end{figure*}

{\it Chiral fluids with odd viscosity}---%
The presence of chiral forces can induce waves in a fluid through a phenomenon known as odd viscosity, which accounts for transverse responses to velocity gradients.
If a fluid is globally rotated as a whole around a fixed axis, anisotropic wave turbulence is typically observed at large scales, while isotropic strong turbulence is observed at small scales~\cite{Zhou1995,Zeman1994,Alexakis2018,Yarom2014,Shaltiel2024}, see Fig.~\ref{fig:intro}a.
By contrast, we find that in odd viscous fluids, wave turbulence is sustained all the way down to the smallest active scales of the flow (Fig.~\ref{fig:intro}b). 
Intuitively, this effect occurs because odd viscosity acts on velocity gradients. Unlike for chiral body forces like Coriolis forces, the generation of waves by odd viscosity is thus enhanced as wavenumber increases, suppressing the transition to strong turbulence. 
Strong turbulence can occur in odd fluids (Fig.~\ref{fig:intro}c), but it is a distinct regime (studied in Refs.~\cite{dewit2024,chen2024}) where the assumption of weakly interacting waves breaks down.
In addition to two dimensional magnetized graphene~\cite{Berdyugin2019} and spinning colloids~\cite{Soni2019}, odd viscosities have been measured in multiple three-dimensional magnetized polyatomic gases as a function of external parameters such as pressure and magnetic field \cite{Beenakker1970}. 
Strong experimental evidence for odd viscosity exists in 3D magnetized plasma (under the name \enquote{gyroviscosity}).
For instance, Refs.~\cite{Stacey2006,Bae2013} discuss experimental signatures of odd viscosity on measurements of plasma discharges in the DIII-D tokamak and Ref.~\cite{Kono2015} on measurements of the dispersion relation of a pair-ion plasma.
All of these systems can be driven or naturally are in a turbulent regime~\cite{Galtier2009,Brizard2007,Terry2000}.

Here, we consider a model chiral fluid that 
bypasses most of the complexities of the aforementioned systems, but 
allows for an exact analytical determination of the turbulent spectra as a function of odd viscosity.
We describe an incompressible chiral fluid with cylindrical symmetry in a direction $\bm{e}_\parallel$ \cite{Khain2022} by the Navier-Stokes equations with odd viscosity
\begin{equation} \label{eq:NS}
     \textrm{D}_t \bm{v} = - \bm{\nabla} p + 
     \begin{pmatrix}
         \nu & \nu_\textrm{odd} & 0 \\
         -\nu_\textrm{odd} & \nu & 0 \\
         0 & 0 & \nu
     \end{pmatrix}
     \Delta \bm{v}
+ \bm{f},
\end{equation}
for the velocity field $\bm{v}$ with $\bm{\nabla} \cdot \bm{v} = 0$, in which $\textrm{D}_t \bm{v} \equiv \partial_t \bm{v} + \bm{v} \cdot \bm{\nabla} \bm{v}$, $p$ is the reduced pressure, $\bm{f}$ the driving force of the flow, $\nu$ the regular viscosity, and $\nu_\textrm{odd}$ is odd viscosity.
For instance, this could arise by having all the particles that compose the fluid to be spinning about $\bm{e}_\parallel$.
The matrix in Eq.~\eqref{eq:NS} is not symmetric when $\nu_\textrm{odd} \neq 0$, manifesting the non-reciprocal character of odd viscosity.
It can be seen as a wavenumber-dependent Coriolis force $-\nu_{\text{odd}}  k^2 \bm{e}_\parallel \times \bm{v}$ that is stronger at large wavevector $k\equiv|\bm{k}|$, as manifested in the dispersion relation \mbox{$\omega_{\bm{k}} = \pm \nu_\textrm{odd} k_\parallel k$} of the odd waves it induces~\cite{Avron1998,dewit2024,chen2024}.
(We have decomposed the wavevector $\bm{k} \equiv (k_\perp \cos\theta, k_\perp \sin \theta, k_\parallel)$ in the basis $(\bm{e}_\perp^1,\bm{e}_\perp^2,\bm{e}_\parallel)$ in which the matrix in Eq.~\eqref{eq:NS} is expressed.)

{\it Weak and strong turbulence}---%
In order to assess whether weak wave turbulence may arise, we must compare the time scale of non-linear interactions $\tau_\textrm{NL}$ with the period of the linear waves $\tau_\textrm{lin}$. 
Here, we estimate $\tau_\textrm{NL}$ as the eddy turnover time, which can be expressed in terms of the kinetic energy spectrum $E(k_\perp,k_\parallel)$ as \mbox{$\tau_\textrm{NL}^{-1} \sim k \sqrt{E(k_\perp,k_\parallel) k_\perp k_\parallel}$}, while \mbox{$\tau_\textrm{lin}^{-1} \sim \omega_{\bm{k}}$} is given by the dispersion relation. The ratio of these timescales is
\begin{equation}\label{eq:chi}
    \chi \equiv \frac{\tau_{\textrm{lin}} } { \tau_{\textrm{NL}} } 
    = \frac{ \sqrt{E(k_\perp,k_\parallel) k_\perp k_\parallel}} {\nu_\textrm{odd} k_\parallel }.
\end{equation}
While regular, eddy-dominated, \textit{strong} turbulence is typically obtained when $\chi \ge \mathcal{O}(1)$, \textit{weak} wave turbulence can emerge when $\chi \ll \mathcal{O}(1)$. 

Which of these regimes of turbulence is encountered in odd viscous flow depends on the scale at which energy is injected. When energy is injected at large scales \mbox{$k_\textrm{in}<k_\textrm{odd}\equiv \varepsilon^{1/4}\nu_\textrm{odd}^{-3/4}$} where $\varepsilon$ is the turbulent energy injection rate \cite{dewit2024}, we encounter Kolmogorov-like turbulence in the range $k<k_\textrm{odd}$ where waves are slow ($\chi > \mathcal{O}(1)$). When the crossover scale $k_\textrm{odd}$ is passed, by construction, the eddy timescale and wave timescale become of the same order and we enter a state which in wave turbulence is known as critical balance \cite{Higdon1984,GS95,Nazarenko2011,Oughton2020,Zhou2021,Alexakis2018} in the range $k>k_\textrm{odd}$ where $\chi\sim\mathcal{O}(1)$, see Fig.~\ref{fig:intro}c. In this range, earlier work has revealed energy accumulation leading to wavelength selection and a spectral scaling $E(k)\sim k^{-1}$ \cite{dewit2024, chen2024}.

\begin{figure}[h]
    \centering
    \includegraphics[width=0.85\columnwidth]{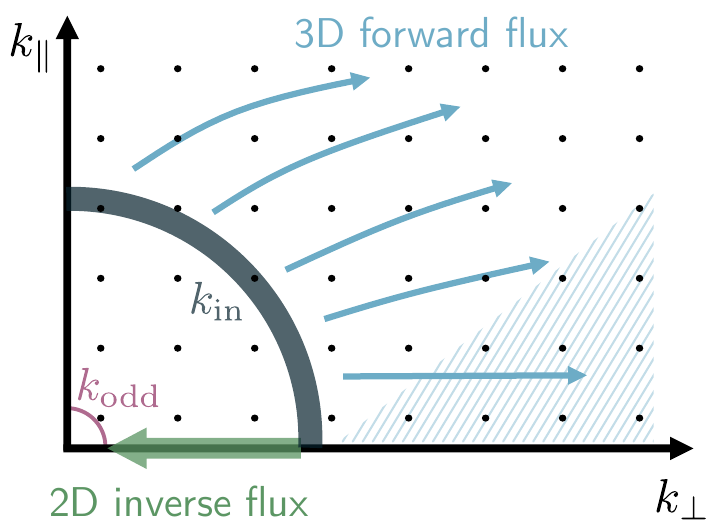}
    \caption{Sketch of the dominant fluxes in the $(k_\perp,k_\parallel)$ space for the weak turbulence case (see Fig~\ref{fig:intro}b). Wave turbulence gives rise to the 3D forward flux depicted here.
    The region $k_\perp \gg k_\parallel$ where the KZ-spectrum is derived is dashed in blue.
    \label{fig:odd_wave_2D_flux_sketch}}
    \vspace{-10mm}
\end{figure}

On the other hand, when injecting energy at smaller scales $k_\textrm{in} > k_\textrm{odd}$, an inverse cascade of kinetic energy is known to emerge in the 2D manifold (with $k_\parallel=0$) in the range $k<k_\textrm{in}$ as detailed in Ref.~\cite{dewit2024}.
However, only a part of the kinetic energy is cascaded upscale through the 2D manifold. A remaining part of the energy flux is still cascaded forward to the range $k>k_\textrm{in}$ (a property also observed in inertial wave turbulence \cite{Shaltiel2024}). This range has not been studied in detail in earlier works. 
As we will show, this is precisely the range where the dynamics is slow enough so that $\chi \ll \mathcal{O}(1)$ for all modes, indicating that we may encounter weak wave turbulence in this regime in the 3D manifold (with $k_\parallel>0$). To enter these flow conditions, we therefore need to force the flow at scales $k_\textrm{in} > k_\textrm{odd}$ and then focus on the range $k > k_\textrm{in}$ (blue region in Fig.~\ref{fig:intro}b, see also Fig.~\ref{fig:odd_wave_2D_flux_sketch} in the End Matter). 

\begin{figure*}[t]
    \centering
    \includegraphics[width=\linewidth]{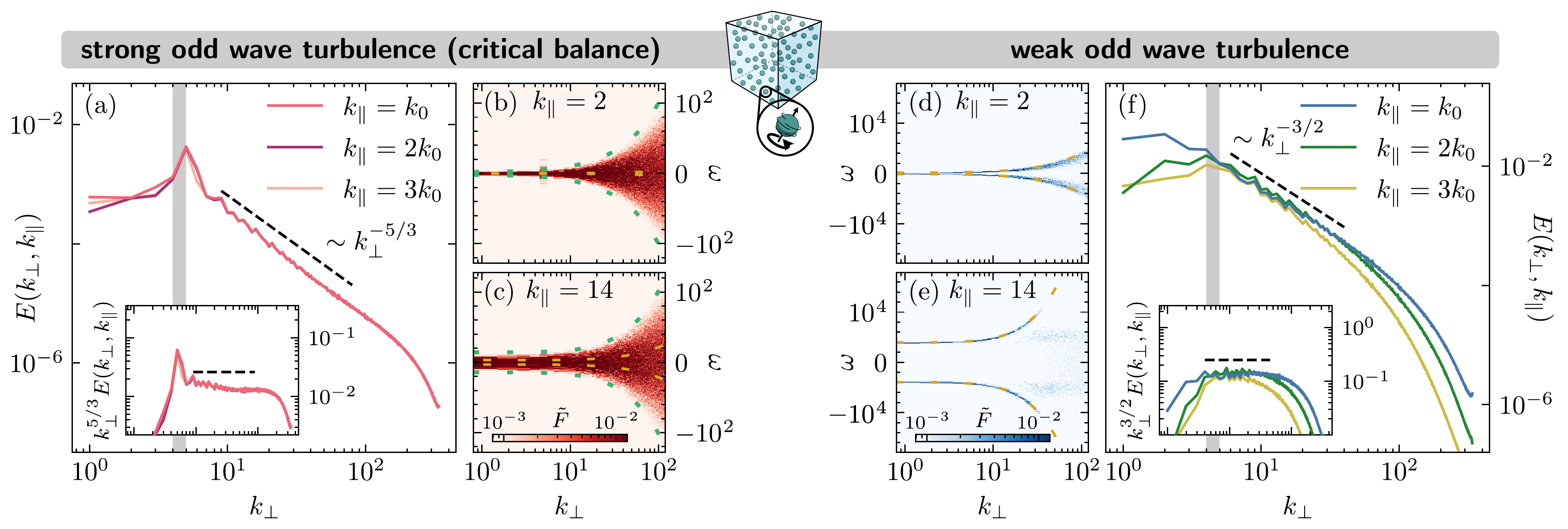}
    \caption{
    \label{odd_viscous_wave_turbulence}
    \textbf{Odd viscous wave turbulence.}
    (a,f) The anisotropic kinetic energy spectra $E(k_\perp,k_\parallel)$ for strong (a) and weak (f) wave turbulence of the first three $k_\parallel$ modes with $k_0 \equiv 2\pi/L_\parallel=1/2$ the smallest wavelength. Insets show the spectra compensated by their respective scaling predictions. The shaded area depicts the forcing range. For the corresponding isotropic energy spectra, see the companion paper \citep{PRE}.
    (b-e) Space-time energy spectra $\tilde{F}(\omega,k_\perp,k_\parallel)$ for the case of strong (b,c) and weak (d,e) wave turbulence, and for different $k_\parallel$ of 2 (b,d) and 14 (c,e).
    Yellow dashed lines indicate the dispersion relation of the odd waves $\omega_k=\pm\nu_\textrm{odd} k_\parallel k$. Green dotted lines in (b,c) indicate the corresponding eddy turnover frequency $\omega\sim k \sqrt{k E(k)} \sim k$ \cite{dewit2024}, which captures the envelope of the space-time spectra for strong turbulence. 
    Spectra are normalized by their corresponding integral over the frequency space for each $k_\perp$, $k_\parallel$, i.e. $\tilde{F}\equiv F(\omega,k_\perp,k_\parallel)/\int F(\omega,k_\perp,k_\parallel) \mathrm{d}\omega$.
    }
\end{figure*}

{\it Weak odd viscous wave turbulence}---%
In the weak regime, one can analytically derive a kinetic equation that governs the energy spectrum, from which the physical properties (e.g. stationary spectra, cascade direction) can be obtained as exact solutions \cite{Nazarenko11,GaltierCUP2023,Newell2011,Zhou2021,Zakharov1992,Benney1966,Zakharov1967,Newell1968}. This calculation is laid out in depth in the companion paper \citep{PRE}. The resulting kinetic energy spectrum, however, can also be obtained from phenomenological arguments.

Weak wave turbulence develops through resonant triadic wave interactions. These interactions happen at time scales much longer than the wave period by a factor $\chi^{-2}$ \cite{GaltierCUP2023,Galtier2023,Zhou2004}.
We can then use a Kolmogorov-type argument and assume that in the weak wave turbulent regime, the energy in each wavenumber shell $E(k_\perp,k_\parallel)k_\perp k_\parallel$ is transported at the wave interaction transfer timescale $\tau_\textrm{tr}\sim\tau_\textrm{lin}/\chi^2$ at a constant energy transfer rate
\begin{equation}
    \varepsilon \sim \frac{E(k_\perp,k_\parallel)k_\perp k_\parallel} {\tau_\textrm{tr}}.
\end{equation}
Using Eq.~\eqref{eq:chi} and assuming $k\sim k_\perp$ (i.e. $k_\parallel \ll k_\perp$), we can then predict the energy spectrum for weak wave turbulence as
\begin{equation}\label{eq:wave_spec}
    E(k_\perp,k_\parallel) \sim \sqrt{\varepsilon \nu_\textrm{odd}} k_\perp^{-3/2} k_\parallel^{-1/2}.
\end{equation}
This yields a timescale ratio \mbox{$\chi \sim \varepsilon^{1/4} \nu_\textrm{odd}^{-3/4} k_\perp^{-1/4} k_\parallel^{-3/4}$}, meaning that the larger $k_\perp$, the weaker the cascade. 
We point out that this is in contrast with rotating (inertial wave) turbulence, where strong turbulence is always recovered at the smallest scales \cite{Galtier2003,NazarenkoS2011}. For this odd wave turbulence, instead, the turbulence remains weak down to the smallest scales, see Fig.~\ref{fig:intro}.

\begin{figure*}[t]
    \centering
    \includegraphics[width=\linewidth]{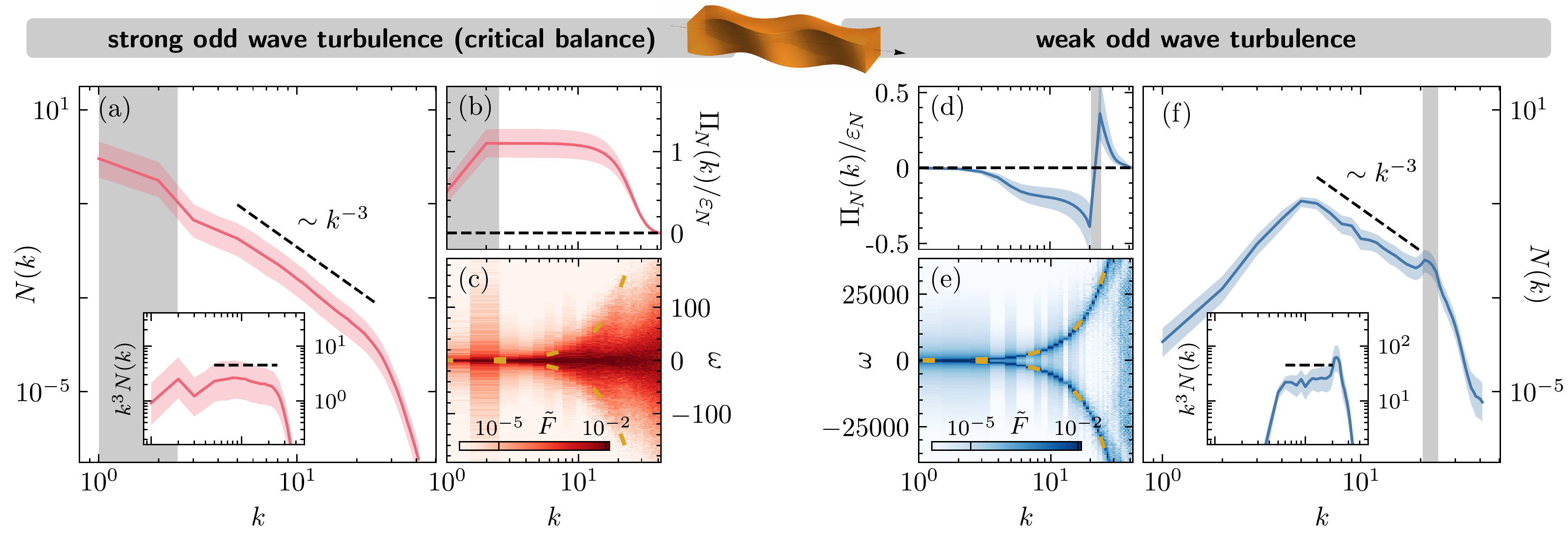}
    \caption{\textbf{Odd elastic wave turbulence.} (a,f) The modal wave action spectra $N(k)$ for the strong critical balance (a) and weak (f) states. Insets show the spectra compensated by their respective scaling predictions. (b,d) The corresponding total flux of modal wave action $\Pi_N(k)$ from wavenumbers $k'< k $ to wavenumbers $k'> k$.
    The shaded gray areas in (a,b and d,f) depict the forcing range. Shaded colored areas depict uncertainty. (c,e) The normalized space-time energy spectra $\tilde{F}(\omega,k)$. Yellow dashed lines indicate the dispersion relation of the odd elastic waves $\omega_k=\pm G_\textrm{odd} k^2$.
\label{fig:odd_elastic_turbulence}
    }
\end{figure*}

{\it Numerical simulations}---%
We numerically solve the 3D Navier-Stokes Eqs.~\eqref{eq:NS} with odd viscosity through direct numerical simulation (DNS) in a periodic box using a pseudo-spectral code (see also Ref.~\cite{dewit2024}). To more closely resolve the limit $k_\perp \gg k_\parallel$, the domain is elongated in the parallel direction $L_\parallel=2L_\perp$, with side length in the perpendicular directions $L_\perp=2\pi$ (results for a cubic domain are provided in the companion paper \citep{PRE}).
The system is forced through Gaussian noise $\bm{f}(\bm{k},t)$ that is delta-correlated in time and is applied in a narrow band of wavenumbers around $k_\textrm{in}=5$, restricted to the 3D manifold. We thus force exactly those wavenumbers with simultaneously $k_\textrm{in} \leq |\bm{k}| < k_\textrm{in}+1$ and $k_\parallel \neq 0$.
To maximize the size of the inertial range, the viscosity term is replaced by a hyperviscosity term of the form $\nu_\alpha \Delta^{\alpha} \bm{v}$. We also introduce a hypoviscous term of the form $\nu_h \Delta^{-\alpha_h} \bm{v}$ to dissipate the inverse flux that develops in the 2D manifold at low wavenumbers.
The input parameters used in the simulations are provided in Tab.~\ref{tab:input} (End Matter).

\begin{figure}[h!]
    \centering
    \includegraphics[width=0.82\linewidth]{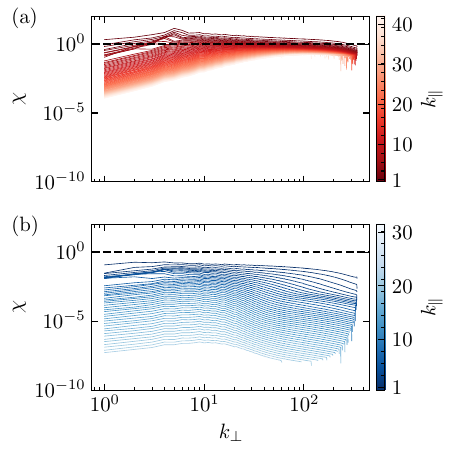}
    \vspace{-3mm}
    \caption{Timescale ratio $\chi \equiv \tau_{\textrm{lin}} / \tau_{\textrm{NL}}$ comparing timescales of odd viscous waves and eddy turbulence for the cases of strong turbulence (a) and weak turbulence (b). Different lines represent different $k_\parallel$. The weak wave turbulence regime is entered when all modes go below $\chi < \mathcal{O}(1)$ (dashed line).}
    \label{fig:chi}
\end{figure}

{\it Numerical results}---%
In order to diagnose whether we have entered the conditions that permit wave turbulence, we assess the timescale ratio $\chi$ in Eq.~\eqref{eq:chi} for all modes in our DNS (Fig.~\ref{fig:chi}, End Matter).
It can be seen that indeed, for the strong turbulence run in Fig.~\ref{fig:chi}(a), modes tend to remain around $\chi \approx 1$, signaling critical balance, while for the weak turbulence run in Fig.~\ref{fig:chi}(b) $\chi \ll 1$ for all modes. There, the slowest mode has $\chi \approx 0.1$, which is in the range where weak turbulence is usually observed in wave-dominated flow systems \cite{Meyrand2018,David2024}. 

To confirm that we have entered the regime of odd wave turbulence, it is crucial to check the spatio-temporal energy spectra 
\cite{Cobelli2009,Yarom2014,LeReun2020,Cosmo2021,Falcon2022,Griffin2022,David2024,Kochurin2024}. To that end, we compute the temporal Fourier transform $\mathcal{F}_t\{...\}$ of time series of selected spatial Fourier modes $\hat{\bm{v}}(k_\perp,k_\parallel,t)$, yielding the spatio-temporal spectrum $ F(\omega,k_\perp,k_\parallel) \equiv \tfrac{1}{2}|\mathcal{F}_t\{\hat{\bm{v}}(k_\perp,k_\parallel,t)\}|^2$.
The results are provided in Fig.~\ref{odd_viscous_wave_turbulence}(b-e). For strong turbulence, this indeed shows a broad range of active modes, its envelope being well captured by the scaling of the eddy turnover frequency $\omega\sim k \sqrt{ k E(k)} \sim k$ \cite{dewit2024}. For weak turbulence, on the other hand, the kinetic energy in the inertial range is very strongly concentrated around the dispersion relation for odd waves $\omega_\textrm{odd}=\pm\nu_\textrm{odd} k_\parallel k$ throughout the inertial range. This is a clear signature that the weak turbulence regime is indeed attained in this case.

Finally, the temporally averaged anisotropic kinetic energy spectra are shown in Fig.~\ref{odd_viscous_wave_turbulence}(a,f).
In the case of strong turbulence, the assumption of critical balance discussed earlier predicts an anisotropic spectrum that scales as \mbox{$E(k_\perp, k_\parallel)\sim k_\perp^{-5/3} k_\parallel^{-1}$} \cite{Higdon1984,GS95,Nazarenko2011,Oughton2020,Zhou2021,Alexakis2018}, which is indeed recovered by the DNS (Fig. \ref{odd_viscous_wave_turbulence}a). 
In the weak turbulence case, we find close agreement with the odd wave turbulence prediction in Eq.~\eqref{eq:wave_spec}, that is $E \sim k_\perp^{-3/2}$ (Fig. \ref{odd_viscous_wave_turbulence}f).
We could not numerically test the scaling prediction for the parallel direction as the inertial range in this direction is too narrow, owing to its very small flux.

{\it Odd elastic wave turbulence}---%
Odd viscous fluids fit in a broader class of odd and non-reciprocal systems \cite{fruchart2026nonreciprocal} which are typically associated with the emergence of waves, and could in principle exhibit wave turbulence. 
In this section, we study as a second example the propagation of one-dimensional nonlinear shear waves in incompressible chiral active solids, where we show that nonlinear transfers across scale take place for emergent conserved quantities associated to the nonequilibrium solid.

In our model, the deformation of the solid is described by the displacement field $\bm{u}(t,z) = [u_x(t,z), u_y(t,z)]$ ($z$ is the direction of propagation) which follows the elastodynamics equation $\rho_0 \partial_t^2 \bm{u} + \Gamma \partial_t \bm{u} = \bm{\nabla}\cdot\bm{\sigma}$ in which $\rho_0$ is the density of the unperturbed medium, $\Gamma$ a linear damping coefficient, and $\bm{\sigma}$ is the stress tensor.
In the overdamped limit where $\rho_0 \omega \ll \Gamma$, this equation reduces to
\begin{equation}
    \label{eom_incomp_soft_solid_overdamped_main}
    \partial_t \bm{u} = G_{\text{odd}} \bm{\epsilon} \partial_z^2 \bm{u} + G^{\text{NL}}_{\text{odd}} 
    \bm{\epsilon} \partial_z\big[\lVert \partial_z \bm{u} \rVert^2 \partial_z \bm{u} \big] + \bm{\mathcal{D}} + \bm{f} ,
\end{equation}
where the linear and nonlinear odd elastic moduli $G_{\text{odd}}$ and $G^{\text{NL}}_{\text{odd}}$ induce a chiral coupling between the two polarizations $u_x$ and $u_y$ \cite{oddreview}, $\bm{\epsilon}$ is the fully antisymmetric matrix (Levi-Civita symbol), $\bm{f}$ is a random forcing, and $\bm{\mathcal{D}}$ is a dissipation term which we assume to be relevant only at very small and/or very large scales
(see the companion paper \citep{PRE} for derivations).
Note that elastic energy is not conserved by Eq.~\eqref{eom_incomp_soft_solid_overdamped_main}: this is due to both the linear damping $\Gamma$ and the odd elastic moduli $G_{\text{odd}}$ and $G^{\text{NL}}_{\text{odd}}$. However, it turns out that the modal wave action $\lVert \bm{u} \rVert^2$ is conserved, as well as another quantity which we call the odd energy. As we now show, when odd elasticity is sufficiently large with respect to normal elasticity, this emergent conserved quantity can be injected through the forcing $\bm{f}$, cascade through scales in the equivalent of an inertial range, and be dissipated at small or large scales. Note that $\bm{\mathcal{D}}$ must dissipate wave action rather than energy, which involves different physical phenomena than usual dissipation. For our simulations, we focus on higher-gradient elastic terms of the form $\tilde{G}_n \partial^{2n}/\partial z^{2n}$ where $\tilde{G}_n$ are known as weakly non-local elastic moduli. In terms of energy, these terms associated to structural dispersion would be conservative, and typically arise in phononic crystals and metamaterials~\cite{Yves2025}. 
Note that when only standard shear elasticity ($n=1$) is present, there is no inertial range as dissipation is present over all scales.

\begin{figure}[b!]
    \centering
    \includegraphics[width=0.9\linewidth]{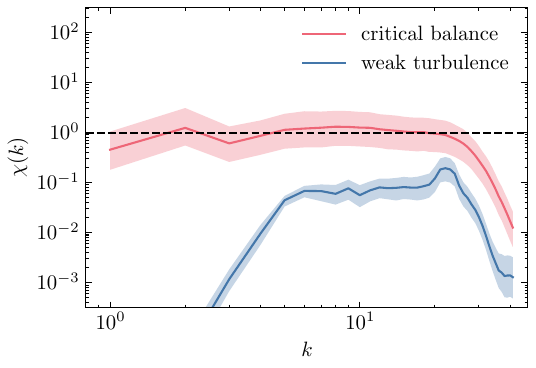}
    \vspace{-3mm}
    \caption{Timescale ratio $\chi \equiv \tau_{\textrm{lin}} / \tau_{\textrm{NL}}$ comparing timescales of odd elastic waves and the non-linearity for the strong critical balance regime (red) and for the weak regime (blue).
    Note that a single line is computed (in contrast with Fig.~\ref{fig:chi}) because the system is 1D.
    }
    \label{fig:chi_elastic}
\end{figure}

Equation~\eqref{eom_incomp_soft_solid_overdamped_main} shares similarities with the Navier-Stokes equation with odd viscosity \eqref{eq:NS}, and admits odd elastic waves at the linear level with dispersion $\omega_k=\pm G_\textrm{odd} k^2$, but due to the assumptions of Cauchy elasticity and incompressibility, it has a third-order nonlinearity at lowest order, corresponding to four-wave interactions (which are in fact irrelevant in the weak regime, so the system is effectively a six-wave system, see \citep{PRE} for details).
Again, a strong regime and a weak regime can be expected, depending on the timescale ratio $\chi\equiv\tau_\textrm{lin}/\tau_\textrm{NL}$ where now $\tau_\textrm{lin}^{-1}=G_\textrm{odd}k^2$ and $\tau_\textrm{NL}^{-1}=G_\textrm{odd}^\textrm{NL} k^5 N(k)$ with $N(k)$ the wave action spectrum (see also Fig.~\ref{fig:chi_elastic}, End Matter). The full weak wave turbulence calculation leading to the wave kinetic equation for the odd elastic solid, as well as the determination and analysis of its solution, are provided in the companion paper \citep{PRE}. This exact analytical calculation reveals that, in the weak regime, an inverse cascade of wave action develops with an associated wave action spectrum $N(k)\propto k^{-3}$. However, the other non-equilibrium solution, which would correspond to a forward cascade of odd energy, is invalid and cannot be realized. To confirm these findings, we numerically solve Eq.~\eqref{eom_incomp_soft_solid_overdamped_main} as detailed in the companion paper \citep{PRE}. The results are provided in Fig.~\ref{fig:odd_elastic_turbulence}, revealing a strong state with a forward cascade (where the flux of modal wave action $\Pi_N(k)$ is positive) and many excited spatiotemporal modes, which seems to be correctly described by critical balance, as well as a weak state, where wave action is concentrated along the dispersion relation, and where an inverse cascade is observed (where $\Pi_N < 0$), in correspondence with the theoretical prediction.

{\it Outlook}---%
To sum up, we have illustrated the application of wave turbulence to odd fluids and solids, paving the way to the study and control of cascades in nonequilibrium media.
When submitted to a forcing at a specific wavelength, these driven-dissipative systems can exhibit turbulent cascades on top of their nonequilibrium steady-state.
The conserved quantities governing the transfer across scales may be radically changed with respect to passive systems, with new effectively conserved quantities emerging in the nonequilibrium steady-state.
As demonstrated, the standard theory of weak wave turbulence successfully predicts their behavior, and can be confirmed numerically.

\medskip

\begin{acknowledgments}
{\it Acknowledgments}---%
This work is supported by the Netherlands Organization for Scientific Research (NWO) through the use of supercomputer facilities (Snellius) under Grant No. 2023.026. This publication is part of the project “Shaping turbulence with smart particles” with Project No. OCENW.GROOT.2019.031 of the research program Open Competitie ENW XL which is (partly) financed by the Dutch Research Council (NWO) and of the ANR project SCASCADE (grant ANR-25-CE30-5080).
SG is supported by the Simons Foundation (Grant No. 651461, PPC). 
M.F., and V.V acknowledge support from the France Chicago center through a FACCTS grant.
V.V. acknowledges support from the Army Research Office under grant nos. W911NF-22-2-0109 and W911NF-23-1-0212, from the National Science Foundation under grant no. DMR-2118415, through the Center for Living Systems (grant no. 2317138) and the National Institute for Theory and Mathematics in Biology (NITMB), from the UChicago Materials Research Science and Engineering Center (NSF DMR-2011864) and from the Theory in Biology program of the Chan Zuckerberg Initiative. 
SG acknowledges V. David for useful discussions. 
\end{acknowledgments}

\bibliography{main}

\clearpage

\onecolumngrid
\begin{center}
    \large \textbf{End Matter}
\end{center}

\begin{table}[h]
\begin{threeparttable}
\caption{\label{tab:input}Input parameters used for the simulations of odd viscous turbulence in this work.
Provided are the box size of the simulation domain in the perpendicular $L_\perp$ and parallel $L_\parallel$ directions, the forcing wavenumber $k_\textrm{in}$, the kinetic energy injection rate $\epsilon$, the odd viscosity $\nu_\textrm{odd}$, the corresponding odd wavenumber $k_\textrm{odd}$, the hypoviscosity $\nu_h$ with power $\alpha_h$, the hyperviscosity $\nu_\alpha$ with power $\alpha$, the integration timestep $\textrm{d}t$ and spatial resolution $N_x \times N_y \times N_z$.}
\begin{ruledtabular}
\begin{tabular*}{0.9\linewidth}{@{\extracolsep{\fill}}lcccccccccccc}
%\toprule
&$L_\perp$&$L_\parallel$&$k_\textrm{in}$&$\epsilon$&$\nu_\textrm{odd}$&$k_\textrm{odd}$&$\nu_h$&$\alpha_h$&$\nu_\alpha$&$\alpha$&$\textrm{d}t$&$N_x\times N_y\times N_z$\\
\midrule
Strong turbulence& $2\pi$ & $4\pi$ & 5 & 0.11 & 0.015 & 13.4 & 0 & N/A & $1.5\times 10^{-14}$ & 3 & $2\times 10^{-5}$ & $1024\times1024\times 256$\\
Weak turbulence& $2\pi$ & $4\pi$ & 5 & 0.11 & 2.0 & 0.34 & 0.2 & 2 & $4.0\times 10^{-15}$ & 3 & $5\times 10^{-6}$ & $1024\times1024\times128$\\
%\bottomrule
\end{tabular*}
\end{ruledtabular}
\end{threeparttable}
\end{table}
\vspace{3mm}

\twocolumngrid

\subsection*{Theory of weak wave turbulence}
The wave turbulence theory is mainly composed of three steps (see the companion paper for detailed derivations \citep{PRE}). 
The first key step is the derivation of the wave amplitude equation.

We first perform a modal expansion of the Navier-Stokes equation to put it in the symbolic form
\begin{equation}
    i \frac{\partial A^{s_k}_k}{\partial t} = s_k \omega_k A^{s_k}_k + M_{kpq}^{s_ks_ps_q} A^{s_p}_p A^{s_q}_q,
\end{equation}
in which summation over $s_p$ and $s_q$ and integration over $\pp$ and $\qq$ are implied (see the companion paper for full expression \citep{PRE}). Here, $A^{s_k}_k$ represents a mode with polarity $s_k = \pm 1$ and wavenumber $k$, and $M_{kpq}^{s_ks_ps_q}$ are nonlinear mode coupling coefficients.
A similar equation would describe any set of waves interacting with quadratic nonlinearities including mode coupling theories of spatially extended quantum systems or optical cavities.
More precisely, \mbox{$ A^{s_k}_k = k \hat \psi_k - s_k k^2 \hat \phi_k$} is defined from the Fourier components of the decomposition \mbox{$\uu = \bnabla \times (\psi \epa) + \bnabla \times (\bnabla \times (\phi \epa) )$} of the velocity field into toroidal ($\psi$) and poloidal ($\phi$) scalar fields ($\hat X_k$ denotes the Fourier transform of a field $X$).
We can show that $\vert A^+_k \vert^2 + \vert A^-_k \vert^2 = 2 \vert \hat \uu_k \vert^2$.

Introducing the interaction representation for weak amplitude waves ($0 < \epsilon \ll 1$) $A_k^{s_k} = \epsilon a_k^{s_k} e^{-is_k\omega_k t}$, we find the wave amplitude equation 
\begin{equation}
\label{WaveEq}
\frac{\partial a^{s_k}_k}{\partial t} = \epsilon L_{kpq}^{s_ks_ps_q} a_p^{s_p} a_q^{s_q} e^{i \Omega_{k,pq} t} \delta_{k,p+q}, 
\end{equation}
in which summation over $s_p$ and $s_q$ and integration over $\pp$ and $\qq$ are implied, where \mbox{$\Omega_{k,p+q} \equiv s_k \omega_k - s_p \omega_p - s_q \omega_q$}, and where $L_{kpq}^{s_ks_ps_q}$ is an interaction coefficient reads (in the anisotropic limit, ie. $k_\perp \gg k_\parallel$)
\begin{equation}
    L_{kpq}^{ss_ps_q} \equiv \frac{\epa \cdot (\ppnb \times \qqnb)}{8 \kpn \ppn \qpn} ( s_p \ppn - s_q \qpn ) ( s \kpn + s_p \ppn + s_q \qpn) .
\end{equation}
The wave amplitude equation describes the slow evolution of odd waves of weak amplitude. We note that the (quadratic) nonlinear coupling vanishes when the wave vectors $\ppnb$ and $\qqnb$ are collinear, or when the wave numbers $p_\perp$ and $q_\perp$ are equal if their associated directional polarities, $s_p$, $s_q$ respectively, are also equal. Interestingly, these properties are also found for inertial wave turbulence \citep{Galtier2003,Monsalve2020}, and more generally for helical waves \citep{Kraichnan1973,Waleffe1992,Turner2000,Galtier2006,Galtier2014}.

Therefore, the long-time statistical behavior is governed by the resonance condition for three-wave interactions, $s_k \omega_k + s_p \omega_p + s_q \omega_q = 0$ and $\kk + \pp + \qq = 0$. These relationships can be written as follows
\begin{equation} \label{reson}
\frac{s_q q - s_p p}{k_\parallel} = \frac{s_k k - s_q q}{p_\parallel} = \frac{s_p p - s_k k}{q_\parallel} .
\end{equation}
For local interactions, $k \simeq p \simeq q$, we obtain $(s_q -s_p)/\kpa \simeq (s_k - s_q)/p_\parallel \simeq (s_p - s_k)/q_\parallel$, which means that the associated cascade is necessarily anisotropic with a negligible cascade along the parallel direction. This situation is reminiscent of inertial wave turbulence \cite{Galtier2003,Lamriben2011} and kinetic Alfvén wave turbulence \cite{David2024} where the waves are also helical. In the following, we take advantage of this property and consider the anisotropic limit $\kpn \gg \kpa$.

The second key step is the derivation of the kinetic equation. 
Assuming statistically homogeneous and anisotropic turbulence \citep{Galtier2024}, one can use a multiple time scale method to derive a kinetic equation of the form
\begin{equation}
    \partial_t E(\kpn,k_\parallel) = - \partial_{\kpn} \Pi_\perp - \partial_{k_\parallel} \Pi_\parallel,
    \label{kinetic_equation}
\end{equation}
describing the evolution of the energy spectrum $E(\kpn,k_\parallel)$, which is related to the amplitudes in Eq.~\eqref{WaveEq} through $E(\kk) \delta (\kk+\kk') = \langle a^{+}_{k} a^{+}_{k'} \rangle + \langle a^{-}_{k} a^{-}_{k'} \rangle$ in the absence of kinetic helicity, where $\langle . \rangle$ is an ensemble average. 
The quantities $\Pi_\perp$ and $\Pi_\parallel$ represent the energy fluxes in the perpendicular and parallel direction. An explicit version of Eq.~\eqref{kinetic_equation} is given below.
This describes the long-time statistical behavior of the dynamics, which is governed by the resonance conditions Eq.~\eqref{reson} for three-wave interactions. Assuming statistically homogeneous and anisotropic turbulence, and the absence of kinetic helicity, the use of the multiple time scale technique yields
\begin{widetext}
\ba \label{KEani2}
\frac{\partial E_k}{\partial t} &=& \frac{\epsilon^2 k_\parallel}{128 \nu_\textrm{odd}} \sum_{s s_{p} s_{q}} \int_{\Delta_\perp}
\left( \frac{\sin \theta_k}{\kpn} \right) \frac{1}{\kpn \ppn \qpn}
\left(\frac{s_p \ppn - s_q \qpn}{k_\parallel} \right)^2 ( s \kpn + s_p \ppn + s_q \qpn)^2 \\
&&\mbox{} \times \left[ \omega_k E_p E_q + \omega_p E_k E_q + \omega_q E_k E_p \right] 
\delta (\Omega_{kpq}) \delta(k_\parallel+p_\parallel+q_\parallel) d\ppn d\qpn dp_\parallel dq_\parallel ,  \nonumber
\ea
\end{widetext}
with $E_k \equiv E(\kpn,k_\parallel)$, $\theta_k$ the angle opposite to $\kknb$ in the triangle $\kknb+\ppnb+\qqnb = {\bf 0}$, and $\Delta_\perp$ an integration domain limited to the resonance conditions.
The density spectrum is defined as $e^{s}(\kk) \delta (\kk+\kk') \equiv \langle a^{s}_{k} a^{s}_{k'} \rangle$. 
In the absence of kinetic helicity, $e^+ = e^- \equiv e$ and the energy spectrum becomes \mbox{$E(\kk) = 2e(\kk)$}. 
Expression (\ref{KEani2}) is the kinetic equation for odd wave turbulence. It is an asymptotically exact equation that does not describe the slow mode ($\kpa=0$), which implies strong turbulence.

The third key step is the derivation of physical properties, the most important of which is the KZ spectrum (see the companion paper for the proof \citep{PRE}). This exact solution is obtained by introducing $E_k = A \kpn^{n} k_\parallel^{m}$ into expression (\ref{KEani2}). Using the Zakharov transformation \cite{GaltierCUP2023} and assuming stationarity, we find $n=-3/2$ and $m=-1/2$.

The cascade direction can be obtained by analyzing the sign of the energy flux. Neglecting the flux in the parallel direction, we obtain 
$\partial E_k / \partial t = - \partial \Pi_\perp (\kpn,k_\parallel) / \partial \kpn$. Using the kinetic equation, we find
$\Pi_\perp^{KZ} = \frac{\epsilon^2 A^2}{384 \nu_\textrm{odd}}  \frac{1}{k_\parallel} I_\perp$, with
\ba    
I_\perp &\equiv& \sum_{s s_p s_q} \int_{\Delta_\perp} \sin \theta_k  
\left( s_q \tilde{q}_\perp - s_p \tilde{p}_\perp \right)^2 \left( s + s_p \tilde{p}_\perp + s_q \tilde{q}_\perp \right)^2  \nonumber \\
&& \times \tilde{p}_\perp^{-5/2} \tilde{q}_\perp^{-5/2} \tilde{p}_\parallel^{-1/2} \tilde{q}_\parallel^{-1/2} 
( \tilde{p}_\parallel  \ln \tilde{p}_\perp +  \tilde{q}_\parallel  \ln \tilde{q}_\perp) \nonumber \\
&& \times \left(1+ \tilde{p}_\perp^{5/2} \tilde{p}_\parallel^{3/2} + \tilde{q}_\perp^{5/2} \tilde{q}_\parallel^{3/2} \right) 
\delta \left( s + s_p \tilde{p}_\perp \tilde{p}_\parallel + s_q \tilde{q}_\perp \tilde{q}_\parallel \right) \nonumber \\
 && \times \delta \left( 1 + \tilde{p}_\| + \tilde{q}_\| \right)  
 d \tilde{p}_\perp d \tilde{q}_\perp d \tilde{p}_\| d \tilde{q}_\| ,
\ea
and with $\tilde p_i \equiv p_i/k_i$ and $\tilde q_i \equiv q_i/k_i$ ($i=\perp,\parallel$). 
A numerical evaluation of the sign of $I_\perp$ shows that the perpendicular energy flux is positive, so this energy cascade is direct.

\end{document}

\title{Supplemental Material:\\Odd wave turbulence}

\maketitle

\section{Isotropic energy spectrum} 

\begin{figure*}[ht]
    \centering
    \includegraphics[width=0.85\linewidth]{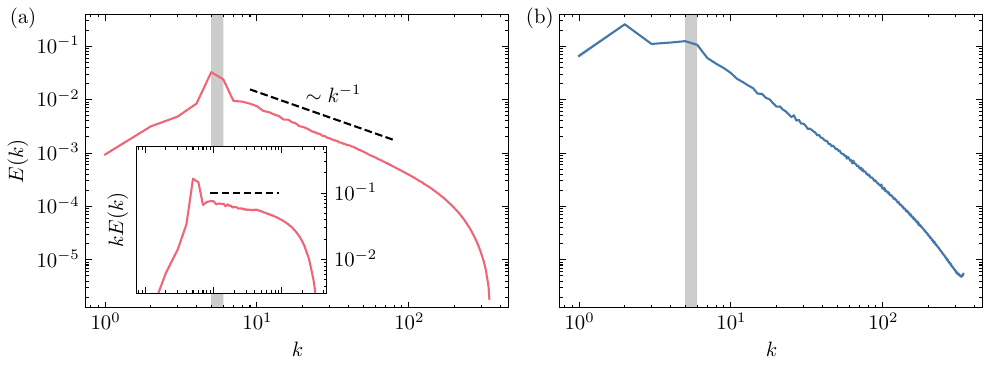}
    \caption{The isotropic kinetic energy spectra $E(k)$ for strong turbulence (a) and weak turbulence (b) as treated in the main text. Inset in (a) shows the spectrum compensated by its respective scaling prediction $E(k)\sim k^{-1}$ as suggested in Ref.~\cite{dewit2024}. For weak turbulence in (b), no scaling prediction exists for the isotropic energy spectrum. The shaded area depicts the forcing range.}
    \label{fig:E_spectra}
\end{figure*}

In Figure \ref{fig:E_spectra}, we present the one-dimensional isotropic energy spectrum from DNS to complement the anisotropic spectrum presented in the main text.

\section{Cubic versus elongated domain}

\begin{figure*}[b]
    \centering
    \includegraphics[width=0.85\linewidth]{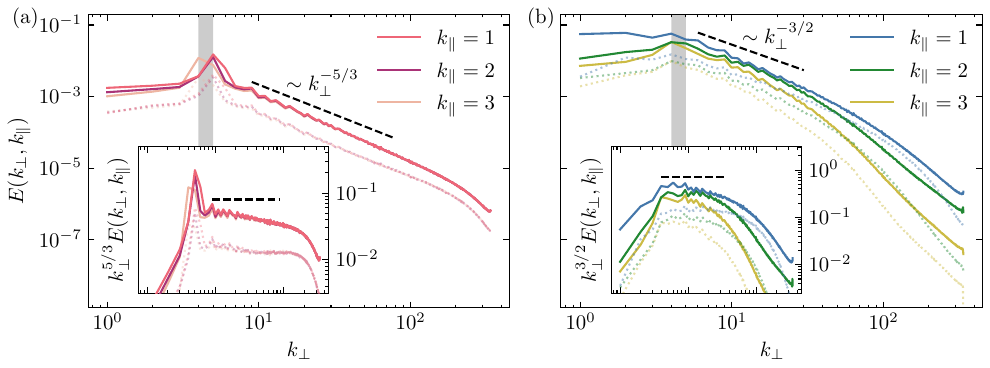}
    \caption{The anistropic kinetic energy spectra $E(k_\perp,k_\parallel)$ for strong turbulence (a) and weak wave turbulence (b) of the first three $k_\parallel$ modes for runs in a cubic domain with $L_\parallel=L_\perp=2\pi$. Insets show the spectra compensated by their respective scaling predictions. The shaded area depicts the forcing range. Dotted lines show the results for the elongated domain with $L_\parallel=2L_\perp$, which is presented in the main text.}
    \label{fig:E_spectra_cube}
\end{figure*}

Considering the anisotropic energy spectrum, for both the regime of strong turbulence in critical balance and that of weak wave turbulence, the derived spectrum is obtained in the limit $k_\perp \gg k_\parallel$. This implies that the numerical agreement should improve if vertically elongated domains are considered, so that smaller wave numbers in $k_\parallel$ are resolved. 
For completeness, we show in Figure \ref{fig:E_spectra_cube} the results of a simulation in a cubic domain with $L_\parallel=L_\perp=2\pi$. We observe that the results agree less well with the theoretical prediction, underpinning the importance of resolving the small wavenumbers in $k_\parallel$. In contrast, for the spectra in $k-\omega$ (not shown), virtually no difference is observable between the elongated domains and the cubic domains.

\bibliography{main}